# Epitaxially Stabilized EuMoO$_3$: A New Itinerant Ferromagnet

Yusuke Kozuka,*,[†] Hidenobu Seki,[†] Takahiro C. Fujita,[†] Suvankar Chakraverty,[‡] Kohei Yoshimatsu,[§] Hiroshi Kumigashira,[⊥,#] Masaharu Oshima,[§] Mohammad S. Bahramy,[‡] Ryotaro Arita,[†,#] and Masashi Kawasaki[†,‡]

[†]Department of Applied Physics and Quantum-Phase Electronics Center (QPEC), University of Tokyo, Tokyo 113-8656, Japan

[‡]Correlated Electron Research Group (CERG), RIKEN-Advanced Science Institute, Wako, Saitama 351-0198, Japan

[§]Department of Applied Chemistry, University of Tokyo, Tokyo 113-8656, Japan

[⊥]Photon Factory and Condensed Matter Research Center, Institute of Materials Structure Science, High Energy Accelerator Research Organization (KEK), Tsukuba, Ibaraki 305-0801, Japan

[#]PRESTO, Japan Science and Technology Agency (JST), Chiyoda-ku, Tokyo 102-0075, Japan

**ABSTRACT:** Synthesizing metastable phase often opens new functions in materials but is a challenging topic. Thin film techniques have advantages to form materials which do not exist in nature since nonequilibrium processes are frequently utilized. In this study, we successfully synthesize epitaxially stabilized new compound of perovskite Eu$^{2+}$Mo$^{4+}$O$_3$ as a thin film form by a pulsed laser deposition. Analogous perovskite SrMoO$_3$ is a highly conducting paramagnetic material, but Eu$^{2+}$ and Mo$^{4+}$ are not compatible in equilibrium and previous study found more stable pyrochlore Eu$_2^{3+}$Mo$_2^{4+}$O$_7$ prefers to form. By using isostructural perovskite substrates, the gain of the interface energy between the film and the substrate stabilizes the matastable EuMoO$_3$ phase. This compound exhibits high conductivity and large magnetic moment, originating from Mo 4$d^2$ electrons and Eu 4$f^7$ electrons, respectively. Our result indicates the epitaxial stabilization is effective not only to stabilize crystallographic structures but also to from a new compound which contains unstable combinations of ionic valences in bulk form.

## ■ INTRODUCTION

Divalent europium compounds have been of considerable interest because the large magnetic moment of 7$\mu_B$ is frequently coupled with electrical or optical properties. As a representative example, EuO exhibits ferromagnetism below a transition temperature of $T_c \sim 70$ K, where an extremely large magnetooptical effect emerges.[1] With doping electrons, EuO additionally shows an insulator-metal transition around $T_c$, accompanied by an extremely large negative magnetoresistance. Complex perovskite oxides composed of EuO and transition-metal (M) oxides, EuMO$_3$,[2] are also under intensive study in recent years to pursue cross-correlated functionalities.[3,4] In particular, EuTiO$_3$, which is an antiferromagnetic insulator in the bulk form, shows a large magnetodielectric effect due to hybridization between Eu$^{2+}$ and Ti$^{4+}$ via oxide ions.[5] This exemplifies the advantage of the perovskite structure for designing new functionalities because separate functions can be embedded to each crystallographic site; magnetism arises from Eu$^{2+}$ ions in A-site, while Ti$^{4+}$ ions in B-site are dielectrically active due to their $d^0$-ness.[6]

Although most of the EuMO$_3$ reported thus far are insulating or semiconducting, a conducting magnetic metal is expected if M is replaced by a transition metal having finite $d$ electrons. In this context, several perovskite oxynitrides such as EuNbO$_2$N, EuTaO$_2$N (Ref. 7), and EuWO$_{1+x}$N$_{2-x}$ (Ref. 8) have been synthesized, which are conducting and show colossal magnetoresistance because of the coupling between electric conduction and the large magnetic moment of Eu$^{2+}$. Within perovskite oxides, EuMoO$_3$ could be one of the promising candidates since analogous SrMoO$_3$ shows the lowest resistivity of $\sim 0.35$ $\mu\Omega$ cm among oxides.[9] However, previous studies were unable to synthesize perovskite Eu$^{2+}$Mo$^{4+}$O$_3$, but resulted in pyrochlore Eu$_2^{3+}$Mo$_2^{4+}$O$_7$.[10] The difficulty in fabricating EuMoO$_3$ is understood based on Ellingham diagram shown in Figure 1, where stable valence states are mapped for simple metal oxides in terms of standard-state Gibbs energy (proportional to oxygen partial pressure) and temperature.[11,12] As seen in Figure 1, Eu$^{2+}$ is stable at low oxygen pressure, while Mo is reduced to be metal in this region. Conversely, when Mo$^{4+}$ is stable at relatively high oxygen pressure, Eu prefers to be trivalent. Here, we report on the successful synthesis of single-crystalline perovskite EuMoO$_3$ in thin film form by pulsed laser deposition (PLD) with using perovskite substrates. Stabilization effect by in-plane lattice matching between the film and the substrate is one of the key factors as has been widely used to synthesize unstable crystallographic structures as reviewed in Ref. 13. However, our result points an important fact that epitaxial stabilization is also effective to form nonequilibrium phases which contain incompatible combination of ionic valence states.

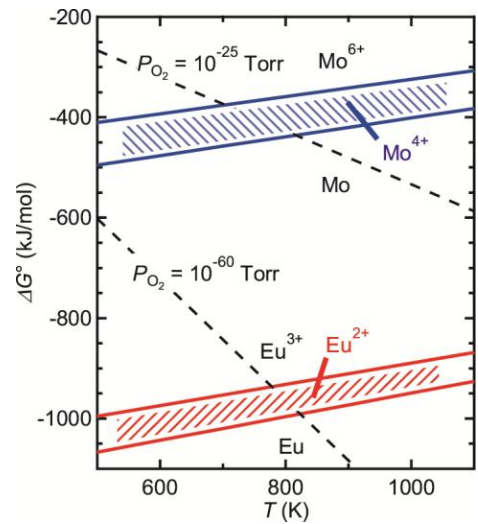

Figure 1. Ellingham diagram for Mo and Eu, which indicate stable conditions of various valence states as oxides. Dashed lines are drawn for two constant pressures.

## ■ EXPERIMENTAL SECTION

**Synthesis.** The films were grown by PLD using a high-density ceramics target with a ratio of Eu/Mo = 1, which was prepared by a hot-press method. As substrates, SrTiO$_3$ (STO) (100) (cubic, 3.905 Å) and GdScO$_3$ (GSO) (110) (orthorhombic, 3.967 Å in pseudocubic setting) were employed. Laser pulses from a KrF excimer laser ($\lambda$ = 248 nm) were focused on the target at a frequency of 5 Hz and a fluence of 2 J/cm$^2$. The chemical composition of the film was confirmed to be Eu/Mo = 1 by energy dispersive X-ray spectroscopy of scanning electron microscope. The distance between the substrate and the target was 5 cm. The crystal structure of the thin films was examined by an X-ray diffractometer (SmartLab, Rigaku).

**Electrical and magnetic characterization.** For electrical measurement, the films were patterned into Hall bar geometry to allow transverse and longitudinal resistivity measurements, which were performed with a Quantum Design PPMS. The temperature and field dependences of the magnetization were measured by a superconducting quantum interference device magnetometer (Quantum Design, MPMS).

**Photoemission spectroscopy.** Photoemission measurements were performed at an undulator beamline of BL2C in Photon Factory, KEK. All spectra were recorded at 300 K using an SES-2002 electron energy analyzer with an energy resolution of about 150 meV. The Fermi Level of the sample was referred to that of gold. The valence band spectra were taken at a photon energy of 600 eV to obtain the highest photon flux in the beam line.

**Band calculation.** The onsite Coulomb-included generalized gradient approximation (GGA + $U$) calculations were carried out using the full-potential augmented plane-wave plus local orbital method, as implemented in the WIEN2K code.[14] The exchange-correlation part of the potential was treated using the Perdew-Burke-Ernzerhof functional[15] with inclusion of onsite Coulomb $U$ and exchange $J$ parameters for Eu 4$f$ states. The muffin-tin radii $R_{MT}$ of Eu, Mo and O atoms were fixed at 2.5, 1.98 and 1.76 Bohr, respectively, and the maximum modulus of the reciprocal vectors $K_{max}$ was defined to give $R_{MT}K_{max}$ = 7. The Brillouin zone was sampled using a 20 × 20 × 20 $k$-mesh.

## ■ RESULTS AND DISCUSSIONS

We first characterized the crystal structure of the films by X-ray diffraction (XRD). Figure 2a shows the result of $\theta$-2$\theta$ measurement for a film grown on STO substrate under optimal growth conditions of 750 °C and 10 mTorr Ar gas containing 3 % H$_2$, which indicates that there are peaks assignable to a perovskite EuMoO$_3$ phase near those of substrate without any of impurity phases. When films were grown under more oxidizing growth conditions such as oxygen or vacuum environment, or lower substrate temperature, Eu$_2$O$_3$ phase is easily formed as detected by the XRD peak around 2$\theta$ = 32° (Figure S1 in Supporting information). Higher growth temperature than 750 °C, on the other hand, results in disappearance of perovskite peaks, which indicate no crystalline phase is formed. Subsequently, the quality of the film was examined by rocking curve of the XRD, which shows a typical full width at half maximum (FWHM) value of 0.2° as shown in Figure 2b. In order to confirm the in-plane relationship between the film and substrate, reciprocal space mapping was measured as shown in Figure 2c, which also support the formation of a perovskite phase with partially relaxed in-plane and out of plane lattice constants of 3.96 Å and 4.00 Å, respectively. It worth noting that these lattice constants roughly matches those of SrMoO$_3$ (Ref. 9) as another evidence of perovskite EuMoO$_3$ phase because of almost identical ionic radii of Sr$^{2+}$ and Eu$^{2+}$ (Ref. 16). For improving the crystal quality, we employed a GSO substrate, which is expected to have smaller lattice mismatch. Surprisingly, we found no film peak in XRD when films were deposited directly on GSO substrate. However, by inserting 5 nm thin STO buffer layer, we could stabilize the EuMoO$_3$ phase (Figure 2d). The film grown on STO-buffered GSO (S-GSO) substrate showed a FWHM value of 0.07° and fully strained lattice constants of 3.97 Å (in-plane) and 3.99 Å (out of plane) (Figures 2e and 2f). The effect of SrTiO$_3$ buffer layer may be explained by the argument of valence-mismatched interface.[17] It is known to have electronic reconstruction when La$^{3+}$Al$^{3+}$O$_3$ is grown on Sr$^{2+}$Ti$^{4+}$O$_3$, yielding in electron doping on Ti site to partially reduce to Ti$^{3+}$. The formation of Eu$^{2+}$Mo$^{4+}$O$_3$ will suffer from energetically unfavorable interface with Gd$^{3+}$Sc$^{3+}$O$_3$, whereas will be able to avoid this with Sr$^{2+}$Ti$^{4+}$O$_3$. Another interface of Sr$^{2+}$Ti$^{4+}$O$_3$/Gd$^{3+}$Sc$^{3+}$O$_3$ may suffer from the same but it will be overcome because SrTiO$_3$ is thermodynamically much more stable than EuMoO$_3$.

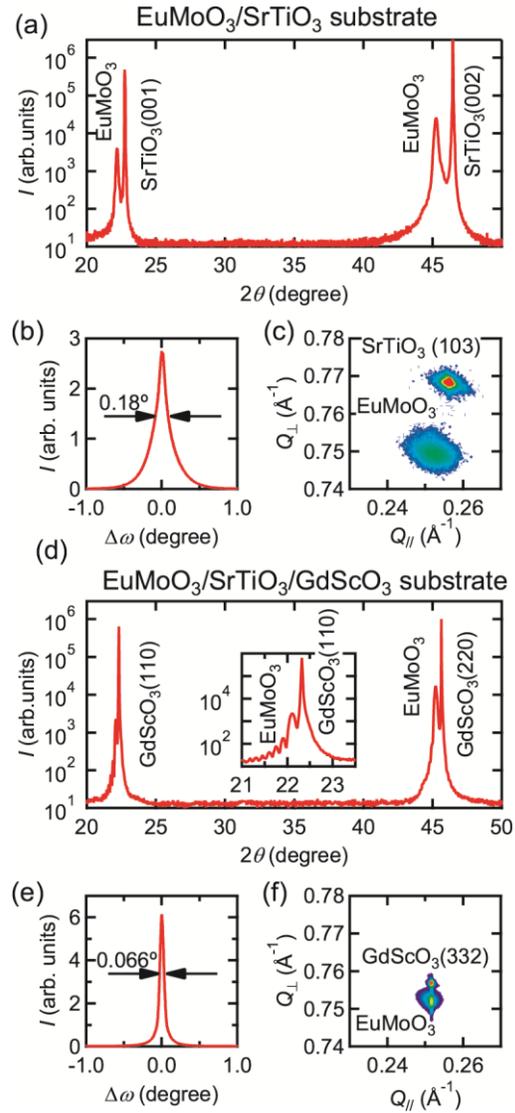

Figure 2. (a) $\theta$-2$\theta$ X-ray diffraction, (b) rocking curve, and (c) reciprocal space mapping around SrTiO$_3$ (103) for EuMoO$_3$ films grown on SrTiO$_3$ substrate. (d)-(f) The same measurements for GdScO$_3$ substrate with SrTiO$_3$ buffer. The reciprocal space mapping for (f) is measured around GdScO$_3$ (332).

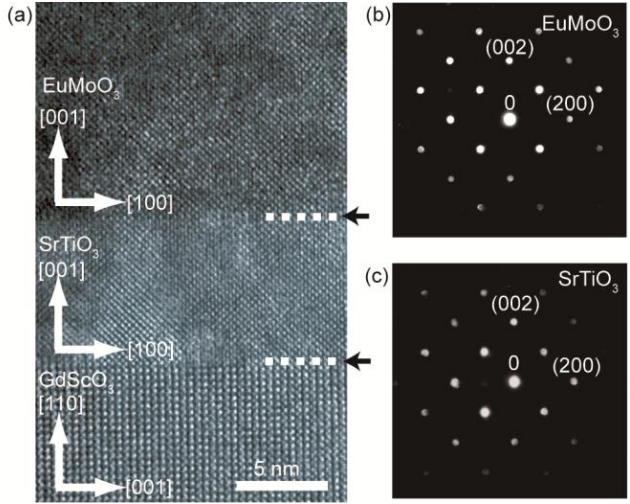

Figure 3. (a) A phase contrast image of high-resolution transmission electron microscopy for a EuMoO$_3$ film grown on GdScO$_3$ substrate with SrTiO$_3$ buffer. Electron diffraction obtained from (b) EuMoO$_3$ film and (c) SrTiO$_3$ buffer layer.

The crystal structure was also examined by a high-resolution transmission electron microscopy (TEM). Figure 3a shows a cross-sectional TEM image with the incidence direction along GSO (1$\bar{1}$0), which gives clear evidence of the epitaxial growth. By comparing the electron diffraction patterns from EuMoO$_3$ film (Figure 3b) with that from STO buffer (Figure 3c), it is evident that EuMoO$_3$ film is isostructural with the perovskite STO. All the above structural measurements evidence the formation of the perovskite EuMoO$_3$ phase. Given the structural information, to confirm the valence states of Eu and Mo ions, we employed X-ray absorption spectroscopy (XAS) and photoemission spectroscopy (PES), respectively. The Eu 3$d$-4$f$ XAS spectrum in Figure 4a indicates Eu$^{2+}$ is dominant, while the amount of Eu$^{3+}$ is negligible.[18] A small shoulder indicating Eu$^{3+}$ may be a result of surface oxidation with possibly compensated by forming Eu$_2$Mo$_2$O$_7$ (Ref. 10) or cation vacancies.[19] The Mo$^{4+}$ was also confirmed by Mo 3$d$ core level spectrum (Figure 4b). These results indicate that perovskite Eu$^{2+}$Mo$^{4+}$O$_3$ is certainly formed.

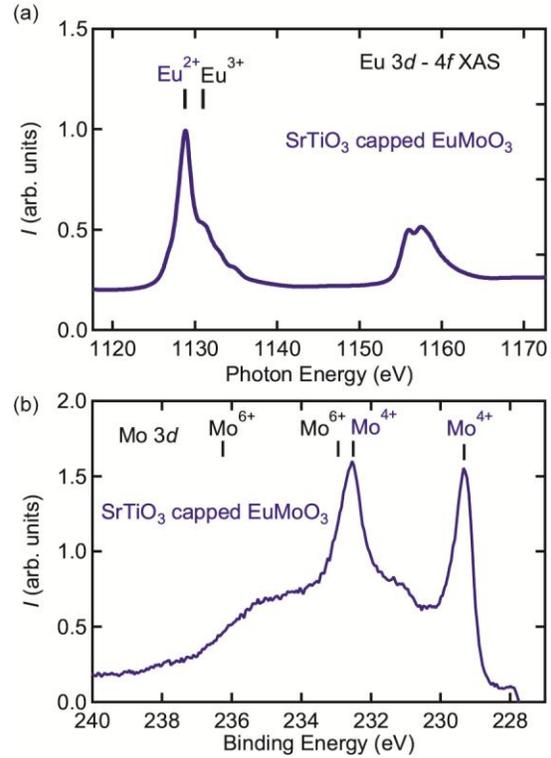

Figure 4. (a) Eu 3$d$-4$f$ X-ray absorption spectra of the EuMoO$_3$ film with SrTiO$_3$ capping layer (2 nm). The spectrum indicates Eu$^{2+}$ is dominant over Eu$^{3+}$ since the absorption around 1129 eV (corresponding to Eu$^{2+}$) is much stronger than that by Eu$^{3+}$ around 1131 eV. (b) Photoemission spectra of Mo 3$d$ core level for the same EuMoO$_3$ film.

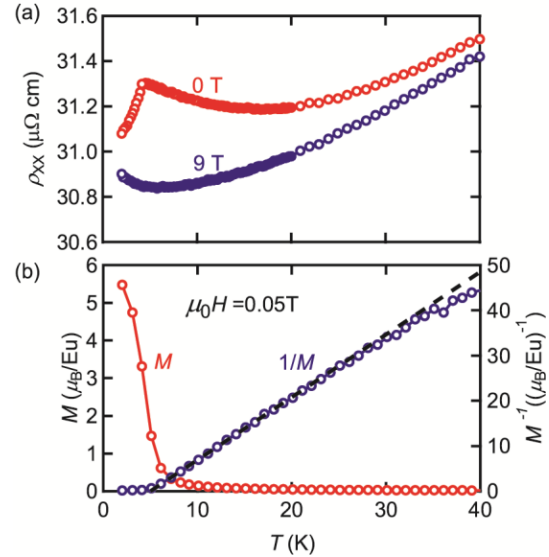

Figure 5. (a) Temperature dependence of resistivity under 0 T and 9 T. (b) Temperature dependence of magnetization measured at 0.05 T. The dashed line is the fit to the 1/$M$ data.

Thus far, we demonstrate the formation of perovskite EuMoO$_3$ thin film in terms of structural and spectroscopic viewpoints. We now present the electrical and magnetic properties of the EuMoO$_3$ thin films. Figure 5a shows temperature de-pendence of resistivity for a film grown on S-GSO substrate, indicating metallic behavior with a residual resistivity of ~ 30 $\mu\Omega$ cm. This value is larger than that of bulk SrMoO$_3$,[9] but similar to that of typical SrMoO$_3$ thin films.[20] From Hall measurement, carrier type was found to be electron and the density was ~ $3 \times 10^{22}$ cm$^{-3}$, which roughly corresponds to 2 electrons/Mo as expected from $d^2$ configuration of Mo$^{4+}$ ion.

Around a temperature of 5 K, we found a kink structure, which is reminiscent of magnetic transition as observed in doped EuTiO$_3$ around $T_c$.[21] With applying a magnetic field, the kink structure disappeared, supporting this scenario. To directly confirm the magnetic transition, we measured magnetization down to 2 K under a magnetic field of 0.05 T as shown in

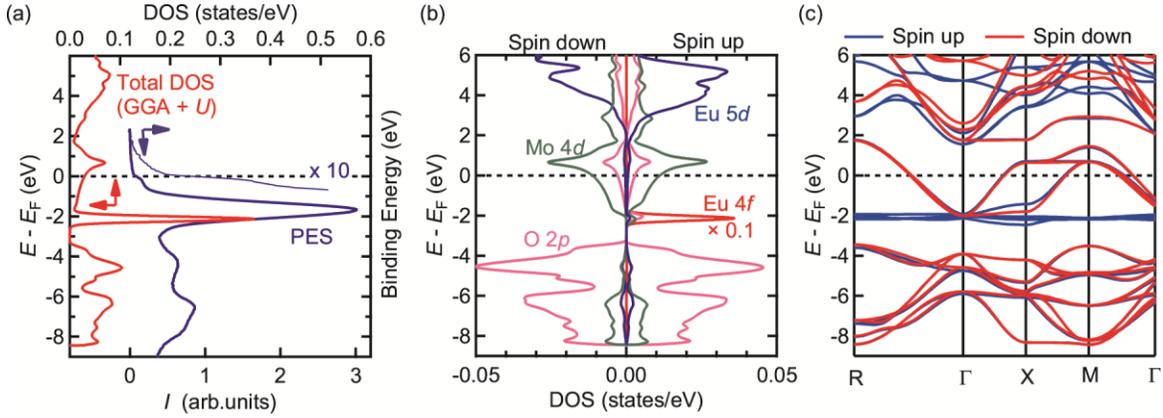

Figure 6. (a) Comparison of density of states calculated by DFT and measured by PES. (b) Density of states for each spin and orbital and (c) band structures calculated by DFT for EuMoO$_3$.

Figure 5b for a EuMoO$_3$ thin film grown on STO substrate. It was not possible to measure magnetization for the film on S-GSO since GSO itself is magnetic.[22] A clear magnetic transition is observed at a Curie temperature of 4.9 K as deduced from a linear fit to the 1/M data, which is close to the kink structure in resistivity.

In order to understand the electrical and magnetic properties, we have performed photoemission measurement as well as a band structure calculation within the limit of GGA + U of density fuctional theory. The valence band spectrum in Figure 6a shows a clear Fermi edge from Mo 4d states, consistent with the metallic conduction, and a sharp peak around -2eV corresponding to the localized Eu 4f bands which contributes to the magnetism. For the sake of comparison, the calculated orbital-projected density of states and spin-polarized band structures are shown in Figures 6b and 6c, respectively. It is to be noted that in order to get the best agreement with PES data, in our calcultions for Eu 4f states, the onsite Coulomb U and exchange J parameters are chosen such that U - J = 7 eV. As can be seen, the calculation also shows that the Mo 4d states most strongly contribute to the bands around the Fermi level in both spin channels, whereas the Eu 4f states form substantially localized bands around -2 eV in the spin-up channel, therefore resulting in a ferromagnetic state with a total magnetic moment of ~ 7.0 $\mu_B$/Eu. Here we note that ferromagnetism is always slightly more favorable (by a few meV difference) than antiferromagnetism, regardless of the value of U - J as shown in Figure S2 of Supporting information for the representative case of U = 0. As expected from the results of PES and the calculation, field-dependent in-plane magnetization in Figure 7a shows that magnetization saturation develops by lowering thetemperature and reaches 6.8 $\mu_B$/Eu. Although the magnetic transition is clear from M-T curve in Figure 5b, we did not observe appreciable hysteresis in the M-H curve, which indicates the coercive field is extremely small.

In contrast to Eu 4f states, Mo 4d states extend widely in energy, crossing the Fermi level, which hybridize with Eu 4f states. This hybridization causes spin polarization in conducting Mo 4d electrons. One of the common ways to detect the spin polarization of conducting electrons is anomalous Hall effect (AHE), as schematically shown in Figure 7c.[23] In ferromagnetic materials, the Hall effect is expressed as $\rho_{xy} = \rho_0 \mu_0 H + \rho_{AHE} M$, where the first term represents the ordinary Hall effect and the second term represents AHE. As shown in Figures 7b and 7d, out-of plane magnetization and the anomalous Hall conductivity, defined by $\sigma_{AHE} = \rho_{xy} / (\rho_{xx}^2 + \rho_{xy}^2)$, becomes nonlinear below $T \sim$ 20 K as approaching the magnetic transition. This consistency between out of plane magnetization and the anomalous Hall conductivity is a proof that the spins of conducting electrons are polarized, possibly through exchange interaction with Eu$^{2+}$ spins.

In this new compound, Eu$^{2+}$ and Mo$^{4+}$ coexist, which is not expected from the Ellingham diagram of simple oxides as an equilibrium phase, where stable phases such as Eu$_2^{3+}$Mo$_2^{4+}$O$_7$ or mixture of EuO and Mo metal prefer to form.[10] This kind of anomalous valence combination is quite rare as a limited example of Ce$^{3+}$V$^{5+}$O$_4$ (Ref. 24). In the example, the lattice energy of the crystal structures stabilizes coexistence of thermodynamically incompatible Ce$^{3+}$ and V$^{5+}$ to allow existence of CeVO$_4$ as a bulk form. In the case of EuMoO$_3$, lattice energy alone cannot stabilize the perovskite phase of EuMoO$_3$. Our result indicates that the epitaxial stabilization can be utilized not only to stabilize crystallographically unstable phases but also to realize nonequilibrium valence combination given the gain of the interfacial energy exceeding the loss of the formation energy in Eu$_2$Mo$_2$O$_7$ → 2EuMoO$_3$ + 1/2O$_2$.

The formation of this new compound brought new characteristics in the perovskite EuMO$_3$ family as it has highly conducting d electrons without doping in contrast to other perovskite europium compounds such as EuTiO$_3$ (Ref. 18) and EuZrO$_3$ (Ref. 25), while the magnetism originates from direct exchange coupling between 4$f$ states of Eu ions, similar to those europium compounds. The coupling between the Mo 5$d$ electrons and Eu 7$f$ magnetic moments lead to spin polarization of the conducting electrons, in contrast to the cases of EuO and EuLiH$_3$, where both conduction and ferromagnetism arises from the Eu sites.[26]

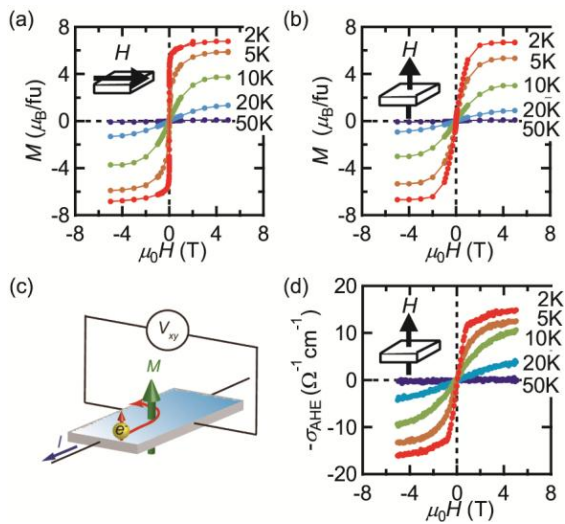

Figure 7. (a) In-plane and (b) out-of-plane magnetization as a function of magnetic field at several temperatures. (c) Schematic diagram of anomalous Hall effect and the measurement geometry. (d) Anomalous Hall conductivity as a function of magnetic field.

## ■ CONCLUSION

In conclusion, we have successfully synthesized perovskite Eu$^{2+}$Mo$^{4+}$O$_3$ thin films on STO and S-GSO by PLD. This compound is found to show high conductivity and large magnetic moment with a magnetic transition temperature of ~ 5 K. The PES and DFT calculation confirmed a ferromagnetic state originating from localized Eu 4$f$ states, while conduction occurs at Mo 4$d$ states. Our finding shed a new light to form compounds containing unusual combinations of valence states using thin film technique. Searching for further new europium perovskite compound incorporating other 4$d$ and 5$d$ transition metal may be intriguing due to strong correlation between electric conduction and large magnetic moment via large spin-orbit coupling.

## ■ AUTHOR INFORMATION
**Corresponding Author**
*E-mail: kozuka@ap.t.u-tokyo.ac.jp

## ■ ACKNOWLEDGMENT
We acknowledge A. Ohkubo (IMR, Tohoku Univ.) for the technical assistance in hot-press, and Y. Tokura for the access to equipment (SQUID). This work was partly supported by "Funding Program for World-Leading Innovative R&D on Science and Technology (FIRST)" Program from the Japan Society for the Promotion of Science (JSPS) initiated by the Council for Science.

# Supporting Information

# Epitaxially Stabilized EuMoO$_3$: A New Itinerant Ferromagnet

Yusuke Kozuka, Hidenobu Seki, Takahiro C. Fujita, Suvankar Chakraverty, Kohei Yoshimatsu, Hiroshi Kumigashira, Masaharu Oshima, Mohammad S. Bahramy, Ryotaro Arita, and Masashi Kawasaki

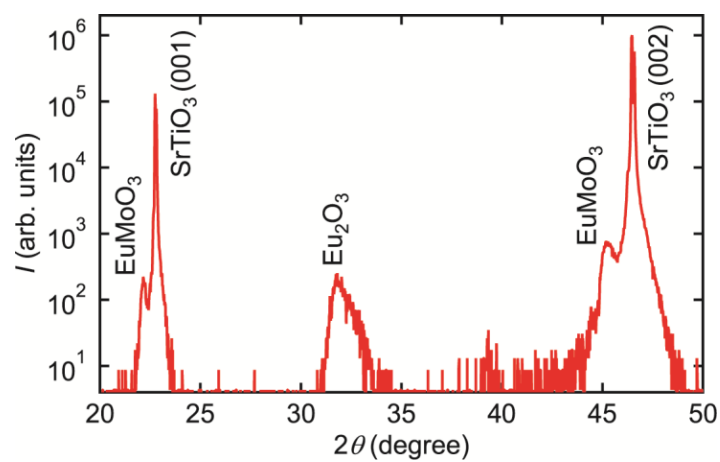

**Figure S1.** $\theta$-2$\theta$ X-ray diffraction pattern for the film grown on SrTiO$_3$ (001) substrate at 650 °C under 10 mTorr Ar + 3 % H$_2$ gas.

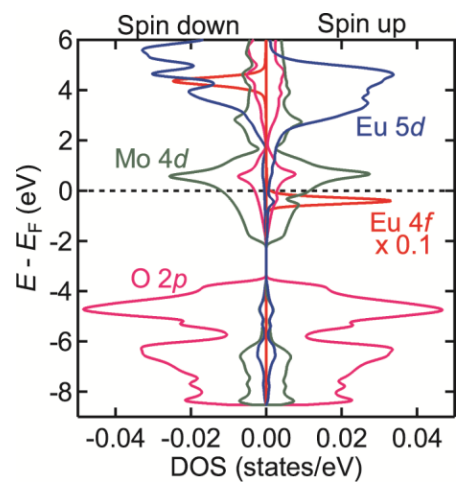

**Figure S2.** Density of states of $EuMoO_3$ for each spin calculated by the generalized gradient approximation calculations without Coulomb interaction $U = 0$